\documentclass{PoS}
\usepackage{psfig,epsfig,colordvi}

\def\lsim{\mathrel{\rlap{\lower4pt\hbox{\hskip1pt$\sim$}}
    \raise1pt\hbox{$<$}}}                
\def\gsi!m{\mathrel{\rlap{\lower4pt\hbox{\hskip1pt$\sim$}}
    \raise1pt\hbox{$>$}}}                

\title{The three-loop $\beta$-function of SU(N) lattice gauge theories with overlap fermions}

\ShortTitle{The 3-loop $\beta$-function for overlap fermions}

\author{\speaker{Martha Constantinou }\\
        Department of Physics, University of Cyprus,\\
P.O.Box 20537, Nicosia CY-1678, Cyprus\\
        E-mail: \email{phpgmc1@ucy.ac.cy}}

\author{Haralambos Panagopoulos\\
        Department of Physics, University of Cyprus,\\
P.O.Box 20537, Nicosia CY-1678, Cyprus\\
        E-mail: \email{haris@ucy.ac.cy}}

\abstract{We briefly report our calculation of the 2-loop coef{f}icient of the coupling constant renormalization function $Z_g$ 
in lattice perturbation theory. The quantity under study is de{f}ined through $g_0=Z_g\,g$,
where $g_0$ ($g$) is the bare (renormalized) coupling constant.
The 2-loop expression for $Z_g$ can be directly related to the 3-loop bare $\beta$-function $\beta_L(g_0)$.

Our calculation is performed using
overlap fermions and Wilson gluons, and the background {f}ield technique 
has been chosen for convenience.
Our results depend explicitly on the number of fermion flavors ($N_f$) and colors ($N$). 
Since the dependence of $Z_g$ on the overlap parameter $\rho$ 
cannot be extracted analytically, we tabulate our results for different values of $\rho$ in the allowed range 
($0<\rho<2$), focusing on values which are being used most frequently in simulations. 
Plots of the 1- and 2-loop results for $Z_g$ versus $\rho$ exhibit a non-trivial dependence on the overlap parameter.

\medskip
\noindent
A longer write-up of this work may be found in Ref.\cite{CP}: \\
\Blue{\tt http://xxx.lanl.gov/abs/0709.4368}}

\FullConference{The XXV International Symposium on Lattice {F}ield Theory\\
		 July 30-4 August 2007\\
		 Regensburg, Germany}

\begin{document}


\section{INTRODUCTION}

In later years, use of non-ultralocal actions which preserve chiral symmetry on the lattice has become more viable.
The two actions which are being used most frequently are overlap fermions
based on the Wilson fermion action and domain-wall fermions.
Overlap fermions~\cite{NN} are notoriously dif{f}icult to study, both numerically and analytically. 
Many recent promising results from simulations with overlap fermions have appeared; see, e.g., Refs.~\cite{IKKSSW}-\cite{BJSSS}. 
Regarding analytical computations, the only ones performed thus far have been either up to 1 loop, such as Refs.~\cite{APV}-\cite{GHPRSS},
or vacuum diagrams at higher loops~\cite{AP,SP}.
{\it{The present work is the {f}irst one involving non-vacuum diagrams beyond the 1-loop level.}}

We compute the 2-loop renormalization $Z_g$ of the bare lattice coupling constant $g_0$ in the presence of overlap fermions.
We relate $g_0$ to the renormalized coupling constant $g_{\rm \overline{MS}}$ as de{f}ined in the ${\rm \overline{MS}}$ scheme
at a scale $\bar{\mu}$; at large momenta, these quantities are related as follows
\begin{equation}
\alpha_{\rm \overline{MS}}(\bar{\mu}) = \alpha_0 + d_1(\bar{\mu} a)\alpha_0^2 + d_2(\bar{\mu} a)\alpha_0^3 + ... \,,
\end{equation}
($\alpha_0=g_0^2/4\pi, \, \alpha_{\rm \overline{MS}}=g_{\rm \overline{MS}}^2/4\pi, \, a:$ lattice spacing).
The coef{f}icient $d_{\,1}$ has been known for a long time; several calculations of $d_2$ 
have also appeared, either in the absence of fermions \cite{LW,AFP}, or using ultra-local fermionic actions \cite{CFPVnew,BPnew}. 
Knowledge of $d_2(\bar{\mu} a)$, along with the 3-loop
$\rm \overline{MS}$-renormalized $\beta$-function allows us to derive the 3-loop bare lattice $\beta$-function,
which dictates the dependence of lattice spacing on $g_0$. 
Ongoing efforts to compute the running coupling from the lattice \cite{Sommer,MTDFGLNS}
have relied on a mixture of perturbative and non-perturbative investigations. As an example, relating 
$\alpha_{\rm \overline{MS}}$ to $\alpha_{\rm SF}$ (SF: Schr\"odinger Functional scheme), 
entails an intermediate passage through the bare lattice coupling and the conversion
from $\alpha_{\rm \overline{MS}}$ to $\alpha_0$ is carried out perturbatively.

The paper is organized as follows:
The theoretical background and the methodology of our perturbative lattice calculations are given in Section II. 
Section III describes the perturbative expansion of the overlap action.
Details on our computation, numerical results and plots of $Z_g$ can be found in Section IV. 


\section{THEORETICAL BACKGROUND}

The renormalized $\beta$-function describes the dependence of the
renormalized coupling constant $g$ on the scale inherent in the renormalization
scheme (chosen to be the $\rm \overline{MS}$ scheme). A bare $\beta$-function is also de{f}ined for the lattice regularization ($\beta_L(g_0)$)
\begin{equation}
\displaystyle \beta(g_{\rm \overline{MS}}) = \phantom{*} \bar{\mu}{dg_{\rm \overline{MS}}\over d\bar{\mu}} \Big|_{a,g_0}, \qquad
\beta_L(g_0)= -a{dg_0\over da} \Big|_{g_{\rm \overline{MS}},\,\bar{\mu}}
\label{beta}
\end{equation}
where $a$ is the lattice spacing, $\bar{\mu}$ the renormalization scale and $g_{\rm \overline{MS}}$ $(g_0)$ the renormalized (bare) coupling constant.
In the asymptotic limit, one can write the expansion of Eq.~(\ref{beta}) in powers of $g_0$
\begin{eqnarray}
\beta_L(g_0) &&=\,\,-b_0 \,g^3_0 -b_1 \,g_0^5 - b_2^{L}\,g_0^7 - ... \\
\beta(g_{\rm \overline{MS}}) &&=\,\,  -b_0 \,g_{\rm \overline{MS}}^3 -b_1 \,g_{\rm \overline{MS}}^5 - b_2\,g_{\rm \overline{MS}}^7 + ... 
\end{eqnarray}
The coef{f}icients $b_0, b_1$ are well-known universal constants (regularization independent); 
$b_i^{L}$ ($i \ge 2 $) (regularization dependent) must be calculated perturbatively.
$\displaystyle \beta_L(g_0)$ and $\displaystyle \beta(g_{\rm \overline{MS}})$ can be related using the renormalization function 
$Z_g$, that is
\vskip -1cm
\begin{equation}
\beta^{L}(g_0) = \left( 1 - g_0^2 \,\,{\partial \ln Z_g^2 \over \partial g_0^2}  \right)^{-1}Z_g \,\,\beta(g_0Z_g^{-1})
\label{beta1}
\end{equation}
Computing $Z_g^2$ to 2 loops
\vskip -.75cm
\begin{eqnarray}
&&Z_g^2(g_0,a\bar{\mu})  = 1 + g_0^2\,(2b_0 \ln(a\bar{\mu})+ l_0) +g_0^4\,(2b_1 \ln(a\bar{\mu})+ l_1) + O(g_0^6) 
\label{Zg}
\end{eqnarray}
\vskip -.35cm
\hskip -.75cm and inserting it in Eq. (\ref{beta1}), allows us to extract the 3-loop coef{f}icient $b_2^L$. 
All quantities in Eq.~(\ref{Zg}), except $l_1$ (for overlap fermions), are known. 
The expansion of  Eq.~(\ref{beta1}) in powers of $g_0^2$ provides the following relation for $b_2^L$ 
\vskip -.35cm
\begin{equation}
b_2^L= b_2 -b_1l_{0}+ b_0 l_{1}
\label{b2lrel}
\end{equation}
\vskip -.35cm
\hskip -.75cm Thus, the evaluation of $b_2^{L}$ requires only a 2-loop lattice calculation, of the quantity $l_{1}$.

The most convenient and economical way to proceed with the calculation of $Z_g(g_0,a\bar{\mu})$ 
is to use the background {f}ield technique, 
in which the following relation is valid
\begin{equation}
Z_A(g_0,a\bar{\mu})   Z_g^2(g_0, a\bar{\mu}) = 1
\end{equation}
where $Z_A$ is the renormalized function of the background {f}ield.
In the lattice version of the background {f}ield technique, the link variable takes the form
\vskip -.35cm
\begin{equation}
U_{\mu}(x)= e^{i a g_0 Q_{\mu}(x)}\cdot e^{i a A_{\mu}(x)}
\label{link}
\end{equation}
\vskip -.3cm
\hskip -.75cm ($Q_{\mu}$: quantum {f}ield, $A_{\mu}$: background {f}ield).
In this framework, instead of calculating $Z_g$, it suf{f}ices to compute $Z_A$.
For the above calculation, we consider the background {f}ield 1PI 2-point function, 
both in the continuum and on the lattice;
for notation and details, the reader can refer to \cite{LW}.
The gauge parameter $\lambda$ must also be renormalized, in order to compare lattice and continuum results
\vskip -.75cm
\begin{eqnarray}
\lambda = Z_Q\,\lambda_0 \, , \qquad Z_Q = 1+ g_0^2\, z_Q^{(1)}+...
\label{Z_Q}
\end{eqnarray}
\vskip -.35cm
\hskip -.75cm ($Z_Q$: renormalization function of the quantum {f}ield). 
{F}inally, $Z_g^2$ takes the form
\vskip -.35cm
\begin{equation}
Z_g^2 = \Big[ 1 + g_0^2\, (\nu_R^{(1)} - \nu^{(1)}) + g_0^4\, (\nu_R^{(2)} - \nu^{(2)}) + 
\lambda_0\, g_0^4\, z_Q^{(1)}\frac{\partial \nu_R^{(1)}}{\partial \lambda} \Big]_{\lambda=\lambda_0}
\label{Zg_expanded}
\end{equation}
($\lambda_0=1$: bare Feynman gauge). Here, the only unknown quantity is $\nu^{(2)}$, 
the 2-loop 1PI lattice Green's function of the background {f}ield; 
its gluonic contribution ($N_f$=0) is known. One can {f}ind the expressions for the remaining, known quantities in Ref.~\cite{CP}.

The fermionic contributions are associated with the diagrams of {F}ig. 1 ($\nu^{(1)}$) and {F}ig. 2 ($\nu^{(2)}$). 
In the present work, $\nu^{(2)}$ is perturbatively calculated for the {f}irst time using overlap fermions and Wilson gluons.
For completeness, we also compute the coef{f}icient $\nu^{(1)}$ and compare it with previous results.
For overlap fermions, the mass counterterm (the {f}illed circle in {F}ig. 2) equals zero, by virtue of the exact chiral symmetry of the overlap action;
consequently, diagrams 19 and 20 both vanish.
Diagrams with infrared divergences become convergent only when grouped together (6+12, 7+11, 8+18, 9+17).\\
\centerline{\psfig{figure=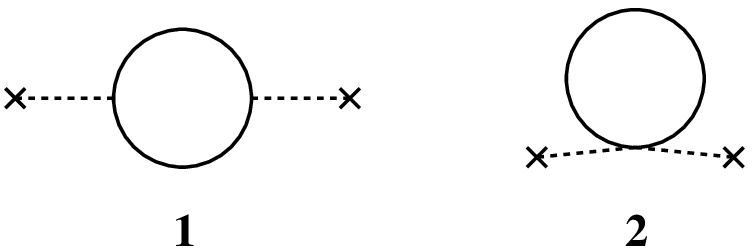,height=1.5truecm}}
{\footnotesize {{\bf {F}ig. 1:} Fermion contributions to the 1-loop function $\nu^{(1)}$.
Dashed lines ending on a cross represent background gluons.
Solid lines represent fermions.}}
\vskip .15cm
\centerline{\psfig{figure=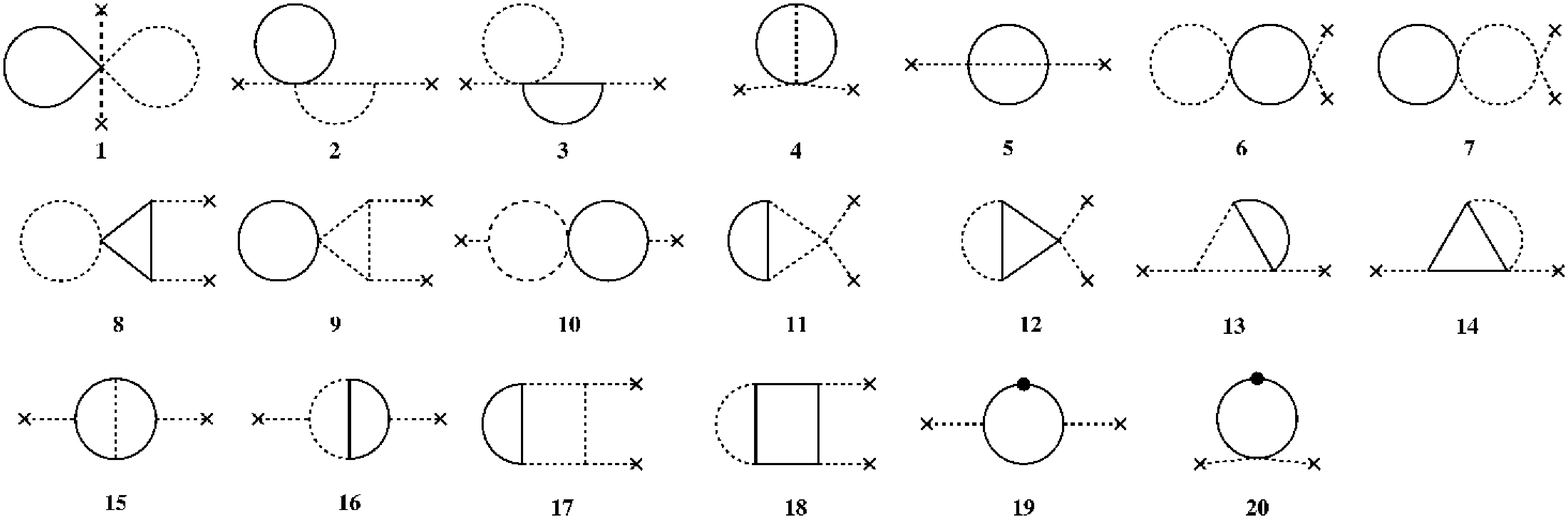,height=5.35truecm}}
\hskip -.6cm
{\footnotesize {{\bf {F}ig. 2:} Fermion contributions to the 2-loop function $\nu^{(2)}$.
Dashed lines represent gluonic {f}ields;
those ending on a cross stand for background gluons.
Solid lines represent fermions.
The {f}illed circle is a 1-loop
fermion mass counterterm.}}


\section{OVERLAP ACTION}

In recent years, overlap fermions are being used ever more extensively in numerical simulations,
both in the quenched approximation and beyond. 
This fact, along with the desirable properties of the overlap action, was our motivation 
to calculate the $\beta$-function with this type of fermions. 
The important advantage of the overlap action is that it preserves chiral symmetry 
while avoiding fermion doubling. It is also ${\cal O}(a)$ improved. 
The main drawback of this action is that it is necessarily non-ultralocal; as a consequence,
both numerical simulations and perturbative calculations are extremely di{f}{f}icult and demanding 
(in terms of human, as well as computer time).

The lattice overlap action is given by
\vskip -.5cm
\begin{equation}
S_{\rm overlap} = a^8 \sum_{n,m} \overline{\Psi}(n) \, D_N (n,m) \, \Psi(m)
\end{equation}
\vskip -.35cm
\hskip -.75cm where $D_N (n,m)$ is the overlap-Dirac operator
\vskip -.75cm
\begin{eqnarray}
D_N (n,m) = \rho \Bigg[\frac{\delta_{n,m}}{a^4}-\left(X\frac{1}{\sqrt{X^\dagger X}}\right)_{nm}\Bigg], \qquad
X = \frac{1}{a^4} \left(D_W -\rho \right)
\end{eqnarray}
\vskip -.25cm
\hskip -.75cm and $D_W$ is the Wilson-Dirac operator.
The overlap parameter $\rho$ is restricted by the condition $0<\rho <2$ to guarantee the correct pole structure of $D_N$.
The coupling constant is included in the link variables, present in the de{f}inition of $X$,
 and one must take the perturbative expansion of $X$ in powers of $g_0$.
This expansion in momentum space takes the form
\vskip -.75cm
\begin{equation}
X(p',p)=\underbrace{\chi_0(p)(2\pi)^4 \delta_P(p' - p)}_{tree-level} + 
\underbrace{X_1(p',p)+X_2(p',p)}_{1-loop} +
\underbrace{X_3(p',p)+X_4(p',p)}_{2-loop} + O(g^5_0)
\end{equation}
\vskip -.25cm
\hskip -.75cm where $\chi_0$ is the inverse fermion propagator and $X_i$ are the vertices of the Wilson fermion action 
($p$ ($p'$): fermion (antifermion) momentum).
Inside of $X_i$'s there appear background and/or quantum gluon {f}ields; their expressions are given in Ref.~\cite{CP}.
Upon substituting the expression for $X_i$'s in the overlap vertices, the latter become extremely lengthy and complicated. 
For instance, the vertex with Q-Q-A-A-$\Psi$-$\overline{\Psi}$ consists of 724,120 terms.

The perturbative expansion of $D_N$ leads to the propagator of zero mass fermions 
and to gluon-fermion-antifermion vertices (with up to 4 gluons for the needs of this calculation).
After laborious analytical manipulations, the overlap-Dirac operator is expanded into terms with up to 4 gluons as
\vskip -.35cm
\begin{equation}
\hskip -0.6cm D_N({k_1},k_2)=D_0(k_1) \,(2\pi)^4\,\delta^4({k_1}-k_2)+{\Sigma({k_1},k_2)}
\label{D_N_detail}
\end{equation}
\vskip -.15cm
\hskip -.75cm $D_0(k_1)$ is the inverse propagator and $\Sigma({k_1},k_2)$ leads to the vertices with up to 4 gluons. 
Due to the large size of $\Sigma({k_1},k_2)$, its explicit form is given in Ref.~\cite{CP}.
%

\section{RESULTS}

For the algebra involving lattice quantities, we make use of our symbolic manipulation package in Mathematica, 
with the inclusion of the additional overlap vertices. The {f}irst step to evaluate the diagrams is the contraction among vertices;
There follow simpl{i}{f}{i}cations of the color dependence, $\rm Dirac$ matrices and tensor structures.
We use symmetries of the theory, or any other additional symmetry that may appear
in particular diagrams, to keep the size of the expression down to a minimum. 
The external momentum $p$ appears in arguments of trigonometric functions and 
the extraction of $p$ dependence is divided into two parts: {f}irst we isolate terms that give single and double logarithms 
(a few thousand terms, expressible in terms of known, tabulated integrals)
and then for the convergent terms we employ naive Taylor expansion with respect to $p$ up to ${\cal O}(p^2)$.
The coe{f}{f}{i}cients of terms proportional to $p^2$ must be numerically integrated over internal momenta. 
The integration is performed by optimized Fortran programs which are generated by our Mathematica `integrator' routine. 
Each integral is evaluated for several lattices with varying lattice size and for different values of the overlap parameter $\rho$.
{F}inally, we extrapolate the results to in{f}inite size lattices; this procedure introduces an inherent systematic error,
which we can estimate quite accurately. Infrared divergent diagrams must be summed up before performing the extrapolation.

We denote the contribution of the $i^{\rm th}$ 1-loop Feynman diagram to $\nu^{(1)}(p)$ as $\nu_i^{(1)}(p)$;
similarly, contributions of 2-loop diagrams to $\nu^{(2)}(p)$ are indicated by $\nu_i^{(2)}(p)$
{\small{
\begin{eqnarray}
&&\widehat{ap}^2 \nu_i^{(1)}(p) = N_f \Bigl[k_{i}^{(0)} + a^2 p^2 \Bigl\{k_{i}^{(1)} + 
k_{i}^{(2)} {\ln a^2 p^2 \over (4 \pi)^2} \Bigr \} + {\cal O}((ap)^4) \Bigr]
\label{nu1Lattice}\\
&&\widehat{ap}^2 \nu_i^{(2)}(p) = N_f \Bigl[c_{i}^{(0)} + a^2 p^2 \Bigl\{c_{i}^{(1)}+ c_{i}^{(2)}\, {\ln a^2 p^2 \over (4 \pi)^2} +
c_{i}^{(3)} \left({\ln a^2 p^2 \over (4 \pi)^2}\right)^2  +c_{i}^{(4)}{\sum_\mu p_\mu^4 \over (p^2)^2} \Bigr\} + {\cal O}((ap)^4)\Bigr]\qquad\quad
\label{nu2Lattice}
\end{eqnarray}
}}
where $\widehat{p}^2 = 4 \sum_\mu \sin^2(p_\mu/2)$. 
The index $i$ runs over diagrams, and the coef{f}icients $k_{i}^{(j)}$, $c_{i}^{(j)}$ depend on the overlap parameter $\rho$.
Moreover, $c_{i}^{(j)}= [ c_{i}^{(j,-1)}/N + c_{i}^{(j,1)}N ]$. These coef{f}icients 
must satisfy conditions coming from comparison with continuum results and usage of Ward Identities; 
we have checked that these conditions are veri{f}ied by our results.
\begin{eqnarray}
&&\bullet\,\, \sum_i k_{i}^{(0)} =0,\,\qquad\,\sum_i c_{i}^{(0)} = 0, \qquad 
\sum_i k_{i}^{(2)} = \frac{2}{3},\,\qquad\,\sum_i c_{i}^{(2)} = {1\over 16 \pi^2} (3 N - {1\over N})\\
&&\bullet\,\, \sum_{i} c_{i}^{(4)}=0, \qquad\,\,
c_{15}^{(3)} = {1\over 3 N}  , \quad c_{16}^{(3)}  = {4\over 3} N  , \quad 
                      c_{17}^{(3)}  = -{5\over 3} N  , \quad c_{18}^{(3)}  = {N^2-1\over 3 N} \qquad\qquad\qquad
\end{eqnarray}

In Table I we tabulate the total 1- and 2-loop coef{f}icients $k^{(1)}\equiv\sum_i k_{i}^{(1)}$, 
$c^{(1,-1)}\equiv\sum_i c_{i}^{(1,-1)}$, $c^{(1,1)}\equiv\sum_i c_{i}^{(1,1)}$
for several values of the overlap parameter (0$<\rho<$2), after extrapolating to $L \rightarrow\infty$. 
The results for $k^{(1)}$  are in good agreement with corresponding results of Ref.~\cite{APV}.
In the Table and all {F}igures, the errors accompanying our results are entirely due to this extrapolation.
In certain cases with large systematic errors we extended the integration up to $L=46$.
In general, the overlap action leads to coef{f}icients which are very small for most values of $\rho$. 
As a consequence, systematic errors, which are by and large rather small, tend to 
be signi{f}icant fractions of the result for $\rho > 1.4$. 
\begin{table}[!h]
\begin{minipage}{17cm}
\label{tab1}
\begin{tabular}{|c|l|l|l|}
 \hline
{\phantom{space}$\rho$\phantom{space}}&\phantom{spacespa}$ {k^{(1)}}$&\phantom{spacespa}$ c^{(1,-1)}$ &\phantom{spacespa}$  c^{(1,1)}$ \\
 \hline
0.2  & \phantom{spac}$$0.01581702(2) \phantom{spac}&	\phantom{spac}-$0.$0044(1) \phantom{spac}& \phantom{spac}\phantom{-}$$0.0118(5) \phantom{spac}\\
0.3  & \phantom{spac}$$0.0133504717(2) \phantom{spac}&  \phantom{spac}	-$0.$00321(6) \phantom{spac} & \phantom{spac}\phantom{-}$$0.0045(1) \phantom{spac}  \\		  
0.4  & \phantom{spac}$$0.0116910952(1) \phantom{spac}& \phantom{spac}  -$0.$00244(4) \phantom{spac} & \phantom{spac}\phantom{-}$$0.0030(1) \phantom{spac}  \\		  
0.5  & \phantom{spac}$$0.0104621922(2) \phantom{spac}& \phantom{spac} 	-$0.$00191(1) \phantom{spac}  & \phantom{spac}\phantom{-}$$0.0022(6) \phantom{spac} \\		  
0.6  & \phantom{spac}$$0.0095058191(2) \phantom{spac}& \phantom{spac} 	-$0.$001606(6) \phantom{spac} & \phantom{spac}\phantom{-}$$0.00176(2) \phantom{spac}  \\
0.7  & \phantom{spac}$$0.00874441051(7) \phantom{spac}& \phantom{spac} -$0.$001397(3) \phantom{spac} & \phantom{spac}\phantom{-}$$0.00145(1) \phantom{spac}  \\		   
0.8  & \phantom{spac}$$0.00813753230(4) \phantom{spac}& \phantom{spac} -$0.$001241(1) \phantom{spac} & \phantom{spac}\phantom{-}$$0.00124(1) \phantom{spac}  \\		   
0.9  & \phantom{spac}$$0.00766516396(3) \phantom{spac}& \phantom{spac} -$0.$001107(1) \phantom{spac} & \phantom{spac}\phantom{-}$$0.001051(9) \phantom{spac}  \\
1.0  & \phantom{spac}$$0.00732057894(3)\phantom{spac}&  \phantom{spac}	-$0.$000979(1)\phantom{spac} & \phantom{spac}\phantom{-}$$0.000872(3)\phantom{spac}  \\		  
1.1  & \phantom{spac}$$0.00710750173(2) \phantom{spac}& \phantom{spac} -$0.$000849(2) \phantom{spac} & \phantom{spac}\phantom{-}$$0.000710(8) \phantom{spac} \\		   
1.2  & \phantom{spac}$$0.00703970232(7) \phantom{spac}& \phantom{spac} -$0.$000706(3) \phantom{spac} & \phantom{spac}\phantom{-}$$0.00052(1) \phantom{spac} \\		   
1.3  & \phantom{spac}$$0.0071425543(2) \phantom{spac}&  \phantom{spac}	-$0.$000543(4) \phantom{spac} & \phantom{spac}\phantom{-}$$0.00033(3) \phantom{spac} \\		  
1.4  & \phantom{spac}$$0.0074569183(2) \phantom{spac}&  \phantom{spac}	-$0.$000335(7) \phantom{spac} & \phantom{spac}\phantom{-}$$0.00007(1)\phantom{spac}  \\		  
1.5  & \phantom{spac}$$0.0080467046(1) \phantom{spac}&  \phantom{spac}	-$0.$00005(1)\phantom{spac} & \phantom{spac}-$0.$0002(1) \phantom{spac}  \\			  
1.6  & \phantom{spac}$$0.0090134204(1) \phantom{spac}&  \phantom{spac}	\phantom{-}$$0.00034(1) \phantom{spac} & \phantom{spac}-$0.$0004(1)\phantom{spac}  \\		  
1.7  & \phantom{spac}$$0.010526080(2)  \phantom{spac}& \phantom{spac}	\phantom{-}$$0.00093(6)  \phantom{spac} & \phantom{spac}-$0.$0021(5) \phantom{spac}\\		  
1.8  & \phantom{spac}$$0.0128914(2)  \phantom{spac}  & \phantom{spac}	\phantom{-}$$0.0020(1)  \phantom{spac}& \phantom{spac}-$0.$02(3) \phantom{spac}    \\
 \hline                  
\end{tabular}
\end{minipage}
\vskip .25cm
\caption{Numerical results for the 1-loop ($\displaystyle {k^{(1)}}\equiv\sum_i k_{i}^{(1)}$) 
and 2-loop ($\displaystyle c^{(1,-1)}\equiv\sum_i c_{i}^{(1,-1)}$, $\displaystyle c^{(1,1)}\equiv\sum_i c_{i}^{(1,1)}$) 
coef{f}icients for different values of the overlap parameter $\rho$.}
\end{table}

We plot the 1-loop coef{f}icient $k^{(1)}$  ({F}ig. 3) and the 2-loop coef{f}icients $c^{(1,-1)}$, $c^{(1,1)}$ ({F}ig. 4)
with respect to $\rho$. The extrapolation errors are visible for $\rho\leq 0.4$ and $\rho\geq 1.7$.\\
\begin{minipage}{0.50\linewidth}
\begin{center}
\psfig{file=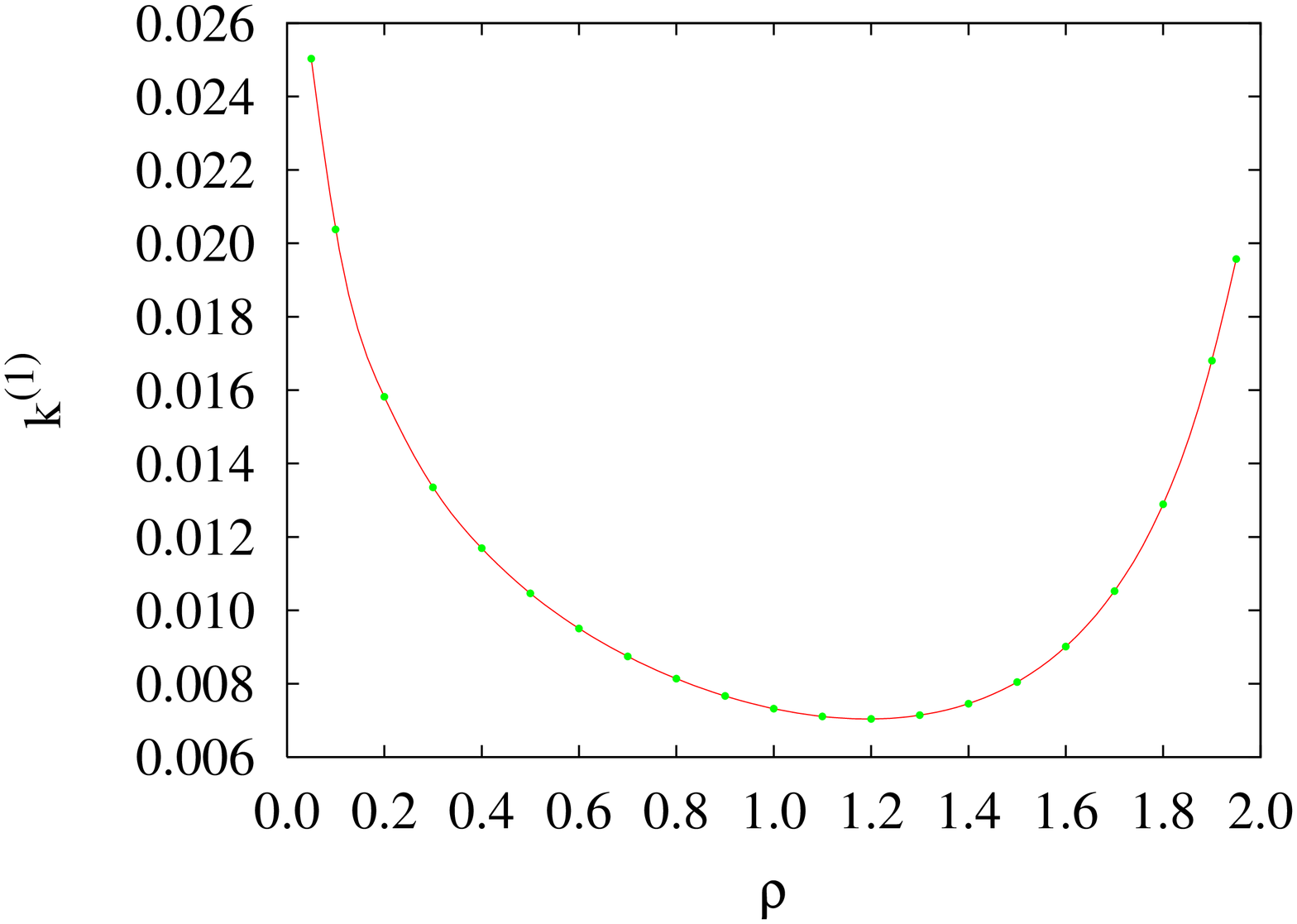,scale=0.31,angle=-90}
{\footnotesize {\noindent{\bf {F}ig. 3:} Plot of the
  total 1-loop coef{f}icient $k^{(1)}$ for different values of the overlap parameter $\rho$.}}
\end{center}
\end{minipage}\hskip0.05\textwidth
\begin{minipage}{0.50\linewidth}
\begin{center}
\phantom{a}\hskip -0.6cm\psfig{file=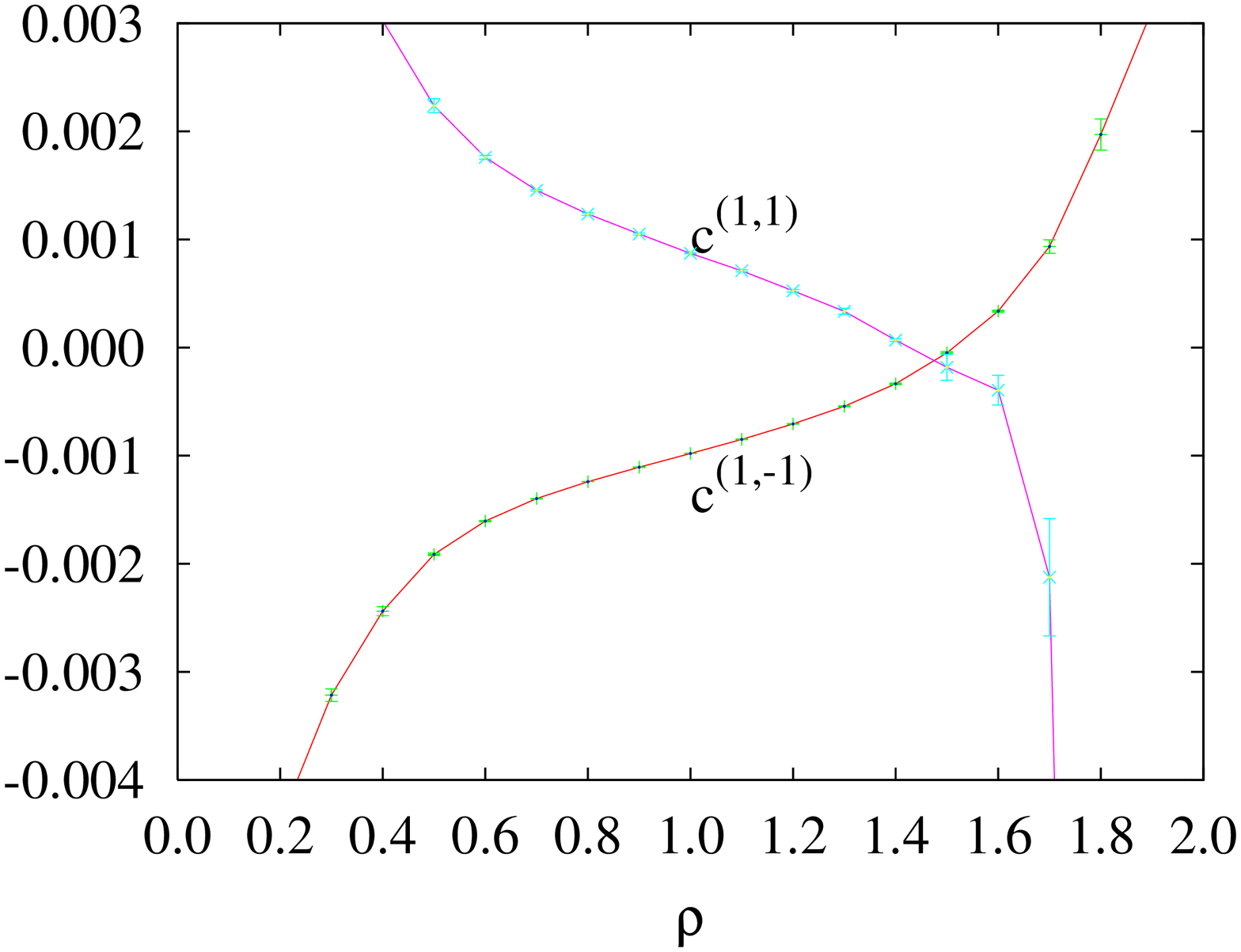,scale=0.31,angle=-90}
\footnotesize {\noindent{\bf {F}ig. 4:} Plot of
  the total 2-loop coef{f}icients $c^{(1,1)}$, $c^{(1,-1)}$ against $\rho$.}
\end{center}
\end{minipage}
\vskip .25cm
\hskip -.75cm {\bf Acknowledgements: } This work is supported in part by the
Research Promotion Foundation of Cyprus (Proposal Nr: $\rm ENI\Sigma X$/0505/45).

\end{document}